\documentclass{article}

\usepackage[preprint]{neurips_2024}

\usepackage{amsmath,amssymb,amsfonts}
\usepackage[utf8]{inputenc} 
\usepackage[T1]{fontenc}    
\usepackage{graphicx}
\usepackage{array}
\usepackage{url}            
\usepackage{booktabs}       
\usepackage{amsfonts}       
\usepackage{nicefrac}       
\usepackage{microtype}      
\usepackage{xcolor}         

\usepackage{float}
\usepackage{hyperref}

\title{PepDoRA: A Unified Peptide Language Model via Weight-Decomposed Low-Rank Adaptation}

\author{
  Leyao Wang,$^{1,*}$ Rishab Pulugurta,$^{1,*}$ Pranay Vure,$^{1,*}$ Yinuo Zhang,$^{1,*}$ Aastha Pal,$^{1}$ \\
  \textbf{Pranam Chatterjee$^{1,2,3,\dag}$} \\\\
  $^{1}$Department of Biomedical Engineering, Duke University \\
  $^{2}$Department of Computer Science, Duke University \\
  $^{3}$Department of Biostatistics and Bioinformatics, Duke University \\\\
  $^{*}$These authors contributed equally \\
  $^{\dag}$Corresponding author: \texttt{pranam.chatterjee@duke.edu}
}

\begin{document}

\maketitle

\begin{abstract}
Peptide therapeutics, including macrocycles, peptide inhibitors, and bioactive linear peptides, play a crucial role in therapeutic development due to their unique physicochemical properties. However, predicting these properties remains challenging. While structure-based models primarily focus on local interactions, language models are capable of capturing global therapeutic properties of both modified and linear peptides. Protein language models like ESM-2, though effective for natural peptides, cannot however encode chemical modifications. Conversely, pre-trained chemical language models excel in representing small molecule properties but are not optimized for peptides. To bridge this gap, we introduce PepDoRA, a unified peptide representation model. Leveraging Weight-Decomposed Low-Rank Adaptation (DoRA), PepDoRA efficiently fine-tunes the ChemBERTa-77M-MLM on a masked language model objective to generate optimized embeddings for downstream property prediction tasks involving both modified and unmodified peptides. By tuning on a diverse and experimentally valid set of 100,000 modified, bioactive, and binding peptides, we show that PepDoRA embeddings capture functional properties of input peptides, enabling the accurate prediction of membrane permeability, non-fouling and hemolysis propensity, and via contrastive learning, target protein-specific binding. Overall, by providing a unified representation for chemically and biologically diverse peptides, PepDoRA serves as a versatile tool for function and activity prediction, facilitating the development of peptide therapeutics across a broad spectrum of applications.

\end{abstract}

\section{Introduction}
\subsection{Background}
Peptide therapeutics, such as macrocycles, peptide agonists, and bioactive linear peptides—including anti-cancer, anti-microbial, anti-viral, and signaling peptides—are gaining prominence in drug development due to their unique chemical properties, such as high specificity, favorable safety profiles, and the ability to target a wide range of biological functions \citep{Wang2022, Zorzi2017, Nauck2021, Akbarian2022}. Despite their potential, predicting the functional properties of these peptides, such as bioactivity, binding affinity, and membrane permeability, remains a significant challenge \citep{Brcenas2022}. This difficulty is particularly pronounced for modified peptides—those incorporating structural alterations like cyclization or unnatural amino acids—which exhibit therapeutic advantages over their natural counterparts, including increased stability and improved bioavailability \citep{Li2024, Zorzi2017, Oeller2023}.

Traditionally, structure-based models have been the primary tools for peptide design. Methods such as FlexPepDock, AlphaFold-based peptide co-folding, and molecular dynamics simulations are commonly used to determine the binding conformation and interactions of peptides with their targets \citep{London2011, Geng2019, Bryant2023}. However, these structure-based methods are generally not well-suited for predicting the broader therapeutic properties of peptides, such as stability, permeability, or bioactivity, as they focus primarily on local interactions rather than capturing the global sequence properties necessary for understanding the full therapeutic potential of peptides \citep{Hoang2021, Zeng2022}. 

Trained on millions of natural amino acid sequences, protein language models (pLMs), such as ESM-2 and ProtT5, have captured important structural and functional relationships inherent in protein sequences \citep{Lin2023, Schmirler2024}. Similar transformer-based methods have been extended to predict the properties of peptides, with models such as PeptideBERT \citep{Guntuboina2023}. In recent work, our lab has successfully used ESM-2 to design linear peptide binders for therapeutic applications, including models such as PepPrCLIP, PepMLM, and SaLT\&PepPr, which generate effective peptide binders from sequence alone \citep{Bhat2023, Chen2024pepmlm, Brixi2023}. Without needing requirement of stable tertiary structure as input, these models enabling therapeutic design of binders and degraders to conformationally diverse targets \citep{Bhat2023, Chen2024pepmlm, Brixi2023}. Nonetheless, models like ESM-2 and PeptideBERT cannot tokenize the structural complexity of modified peptides with cyclization, non-canonical amino acids, or other modifications \citep{Lin2023}. 

Contrastively, pre-trained chemical language models (cLMs), such as ChemBERTa, ChemFormer, and MolT5, are trained primarily on small molecule datasets, using SMILES notations to encode molecular structures for downstream tasks such as property prediction and molecular generation \citep{Ahmad, Irwin2022, Edwards}. Because these models are largely optimized for small molecules, they do not naturally account for the sequential and structural intricacies of peptides, limiting their utility in peptide-centric applications. Recently,  PeptideCLM was pre-trained on peptide-specific SMILES data from scratch, via the RoFormer model architecture, performing strongly on membrane permeabilization prediction \citep{Feller2024}. PeptideCLM, however, was trained solely on modified peptide representations, thus the model may not retain valuable physicochemical information that pre-trained cLMs have already learned from broader molecular datasets, as well as key properties of bioactive linear peptides composed of only wild-type amino acids. 

To overcome these challenges and bridge the gap between natural and modified peptide representation, we introduce PepDoRA—a unified peptide representation model designed for both bioactive linear peptides and structurally modified variants. By efficiently fine-tuning the state-of-the-art ChemBERTa-77M cLM \citep{Ahmad} on both modified and natural peptide sequence data via efficient weight-decomposed low-rank adaptation (DoRA) \citep{DoRA}, PepDoRA generates optimized embeddings capturing core therapeutic properties, including membrane permeability, non-fouling and hemolytic propensity, and most importantly, target-specific binding. In total, PepDoRA serves as a versatile tool for both prediction and design tasks, supporting the development of therapeutic peptides across a broad spectrum of applications.


\section{Methods}

\subsection{Dataset}
The dataset used in this study consists of 100,000 peptides, including both modified and bioactive sequences. The modified peptides were sampled from the pre-training dataset used to train PeptideCLM, incorporating cyclic structures and unnatural amino acids \citep{Feller2024}. Bioactive linear peptides, such as anti-cancer, anti-viral, and anti-microbial peptides, were sourced from the BIOPEP-UWM database \citep{Minkiewicz2019}. Additionally, linear peptide binders to target proteins were obtained from the Propedia v2.3 database \citep{Martins2023}. All peptides were encoded using the SMILES string notation via RDKit \cite{Bento2020}, after which they were shuffled, combined, and then split into 80\% training data and 20\% testing data.

\subsection{Masked Language Modeling (MLM) Task}
The Masked Language Modeling (MLM) task was employed to fine-tune the ChemBERTa-77M-MLM model for peptide representation \citep{Ahmad}. Given a peptide sequence represented as a SMILES string, a subset tokens is randomly selected and masked. The model then learns to predict the masked tokens, based on the surrounding context, to optimize the following objective:

\begin{equation}
    L_{\text{MLM}} = -\mathbb{E}_{(x, \tilde{x}) \sim D} \left[ \sum_{i \in \mathcal{M}} \log P(x_i | \tilde{x}) \right]
\end{equation}

where $x$ represents the original sequence, $\tilde{x}$ is the sequence with masked tokens, $\mathcal{M}$ denotes the set of masked positions, and $P(x_i | \tilde{x})$ is the predicted probability of the masked token $x_i$ given the masked sequence $\tilde{x}$. The model was trained using a standard MLM approach with a masking rate of 15\%.

\subsection{Fine-Tuning of ChemBERTa}
To adapt the ChemBERTa model for peptide prediction tasks, we employed three different fine-tuning strategies: (1) fine-tuning the last layer of ChemBERTa-77M-MLM \citep{Ahmad}, (2) using LoRA \citep{LoRA}, and (3) using DoRA \citep{DoRA}. In both cases, LoRA or DoRA adapters are incorporated into the last three layers, specifically targeting the query, key, and value modules.

\subsubsection{Fine-Tuning the Final Layer}
In the first approach, all weights of the final transformer layer of ChemBERTa-77M-MLM \citep{Ahmad} were unfrozen, including the attention weights, feed-forward weights, and layer normalization parameters, and updated during training. This allowed for focused adaptation of the model while minimizing the number of parameters being optimized.

\subsubsection{Low-Rank Adaptation (LoRA)}
LoRA aims to fine-tune a pre-trained model by adding a low-rank decomposition to the weight update \citep{LoRA}. Specifically, for a pre-trained weight matrix $W_0 \in \mathbb{R}^{d \times k}$, LoRA models the update $\Delta W \in \mathbb{R}^{d \times k}$ as:

\begin{equation}
    W' = W_0 + BA
\end{equation}

where $B \in \mathbb{R}^{d \times r}$ and $A \in \mathbb{R}^{r \times k}$ are trainable low-rank matrices with $r \ll \min(d, k)$. This approach allows for efficient adaptation by only updating a small number of parameters.

\subsubsection{Weight-Decomposed Low-Rank Adaptation (DoRA)}
DoRA decomposes the pre-trained weight into two components: magnitude ($m$) and direction ($V$), which are fine-tuned separately to enhance learning capacity \citep{DoRA}. The decomposition of the weight matrix $W$ is given by:

\begin{equation}
    W = m \frac{V}{||V||_c}
\end{equation}

where $m \in \mathbb{R}^{1 \times k}$ represents the magnitude vector, $V \in \mathbb{R}^{d \times k}$ is the directional matrix, and $|| \cdot ||_c$ denotes the vector-wise norm across each column of $V$. During fine-tuning, the weight update is given by:

\begin{equation}
    W' = m \frac{W_0 + BA}{||W_0 + BA||_c}
\end{equation}

\begin{figure}[h]
  \centering
  \includegraphics[width=0.8\textwidth]{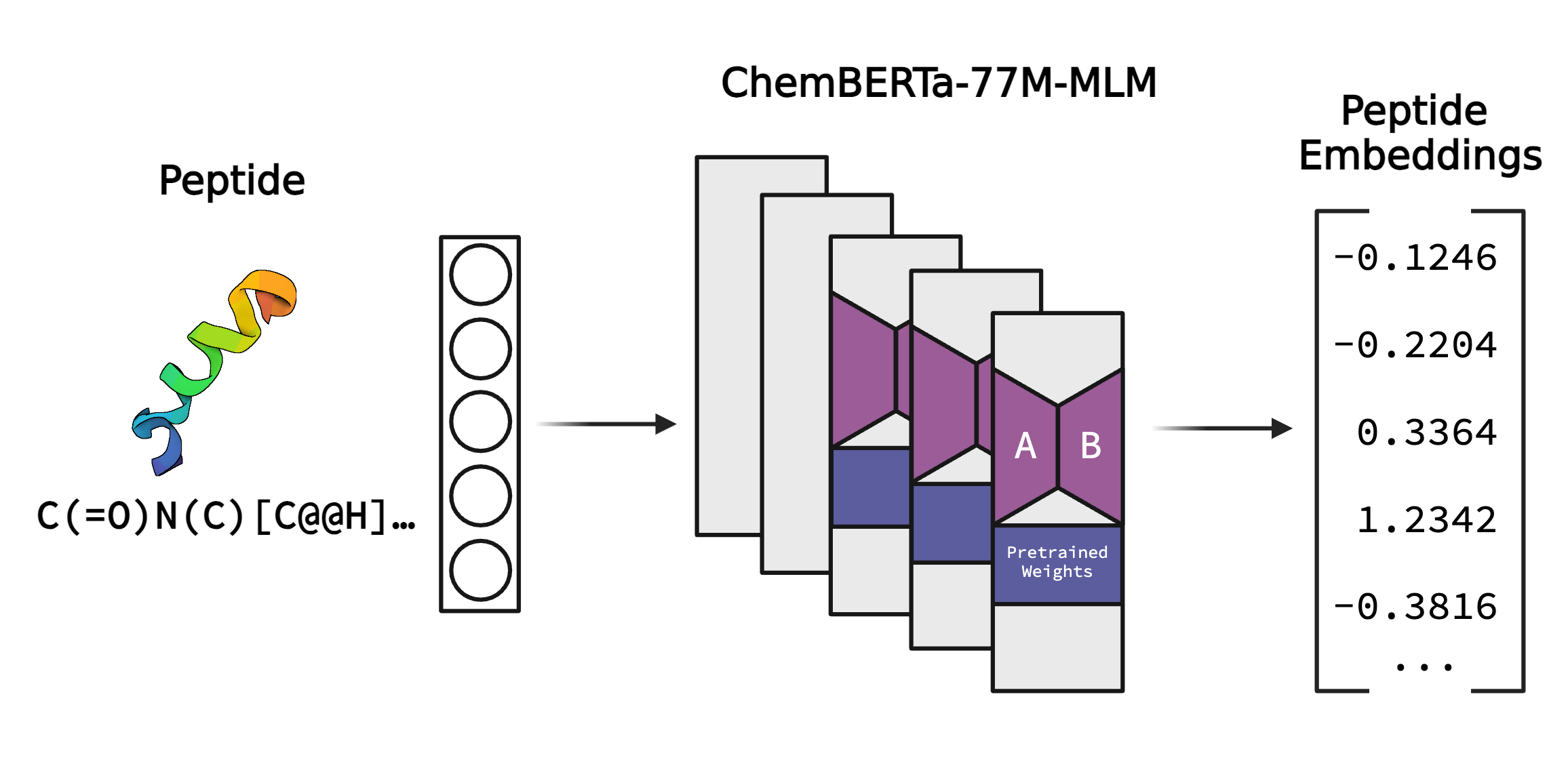}
  \caption{Schematic representation of PepDoRA for peptide embedding generation. The peptide SMILES representation is processed through the ChemBERTa-77M-MLM architecture, with the final three layers fine-tuned via DoRA, resulting in optimized peptide embeddings suitable for downstream prediction tasks.}
  \label{Fig: pepdora_schematic}
\end{figure}

Here, $B$ and $A$ are low-rank matrices similar to LoRA, and $W_0$ is the pre-trained weight. DoRA thus provides an improved adaptation mechanism while maintaining inference efficiency. The A schematic of the resultant PepDoRA is depicted in Figure \ref{Fig: pepdora_schematic}.

\subsection{Model Training}
All models were trained on two NVIDIA A6000 GPUs, each with 48 GB of VRAM. The training of the models was conducted using the \texttt{AdamW} optimizer, with a learning rate of $2 \times 10^{-5}$ and a weight decay of $0.01$. The batch size for training was set to 2, and training was performed for a maximum of 10 epochs. 

\subsection{Evaluation}
\subsubsection{Membrane Permeability Prediction}
The membrane permeability of peptides was predicted using a regression model identical to that described in the PeptideCLM paper \citep{Feller2024}. Specifically, we used a regression approach to predict the logarithm of the diffusion constant for the PAMPA dataset. The dataset was split into six clusters using k-means clustering on the principal components of the peptide embeddings, with the smallest group held out for testing. The remaining data was used in a 5-fold cross-validation setup.

For evaluation, we applied this strategy across the three fine-tuned ChemBERTa models (last layer fine-tuned, LoRA, and DoRA), as well as PeptideCLM-23M and ChemBERTa-77M-MLM. For each model, the language modeling head was replaced with a fully connected feed-forward layer, with a width matching the hidden state of the model, to perform regression for predicting membrane permeability. Model training was conducted for up to 20,000 steps, and checkpoints were saved based on the lowest mean squared error (MSE) observed on the validation set. The final test metrics, including root mean squared error (RMSE), were calculated as the mean of pooled predictions from the five cross-validated models. This process was repeated five times, and the mean across the five iterations were plotted in Figure \ref{Fig: membrane}.

\subsubsection{Non-Fouling and Hemolysis Prediction}

The hemolytic activity and non-fouling property of peptides were predicted using two classification modules. Using the pre-split training datasets curated previously in \cite{Guntuboina2023}, we utilized the XGBoost gradient boosting architecture using tree-based learners, with subsequent hyperparameter optimization via Optuna over 50 trials \citep{Akiba2019}. Model parameters were tuned by optimizing a combined metric incorporating accuracy, precision, recall and F1-score on the validation set.

We evaluated performance across the DoRA-fine-tuned ChemBERTa model, as well as PeptideBERT. PeptideBERT predictions were made using a multilayer perceptron (MLP) with a single fully connected layer of 480 nodes, trained for 30 epochs using the \texttt{AdamW} optimizer (learning rate 0.00001), batch size of 32, and binary cross-entropy loss, as described in their methods. We used the \texttt{ReduceLROnPlateau} scheduler, which reduces the learning rate by a factor of 0.1 if validation accuracy did not improve for four consecutive epochs, identical to the original implementation in \cite{Guntuboina2023}. Both models were evaluated using identical train-validation-test splits to ensure fair comparison.

\subsubsection{Contrastive Language Model for Peptide-Protein Interaction}
To evaluate peptide-protein interactions, we used a contrastive learning model based on our previous PepPrCLIP architecture \citep{Bhat2023}, which itself is based on the contrastive language-image pretraining (CLIP) architecture from OpenAI \citep{CLIP}. The target protein embeddings were generated using ESM-2-650M, while the peptide embeddings were obtained from the three fine-tuned ChemBERTa models, PeptideCLM-23M \citep{Feller2024}, and ChemBERTa-77M-MLM \citep{Ahmad}. The model was trained to maximize the cosine similarity between true peptide-protein pairs and minimize it for non-binding pairs:

\begin{equation}
    L_{\text{CLIP}} = -\frac{1}{n^2} \sum_{i,j} \left[ \log \frac{\exp(\text{sim}(e_i, p_j))}{\sum_{k=1}^n \exp(\text{sim}(e_i, p_k))} \right]
\end{equation}

where $e_i$ and $p_j$ are the embeddings of the target protein and peptide, respectively, and $\text{sim}(\cdot, \cdot)$ represents the cosine similarity. Three metrics were used to evaluate the performance of this model:
\begin{itemize}
    \item \textbf{Binary Accuracy}: This metric measures the percentage of correctly classified peptide-protein pairs, determining whether a given peptide binds to a specific protein or not.
    \item \textbf{Top-1 Accuracy}: This metric evaluates the model's ability to correctly identify the single best peptide-protein pair from all possible candidates. This is particularly important for identifying the most probable binding partner in cases where a direct interaction prediction is needed.
    \item \textbf{Top 10\% Accuracy}: This metric measures whether the true peptide-protein pair is ranked within the top 10\% of all possible pairs. This is particularly useful in scenarios such as high-throughput screening, where identifying a shortlist of potential candidates is of high value.
\end{itemize}
The model was trained on a curated \textasciitilde 12,000 peptide-protein pairs from the Propedia v2.3 \citep{Martins2023} and PepNN \citep{Abdin2022} datasets, which were curated previously \citep{Chen2024pepmlm}. To ensure the most robust evaluation, the entire PepNN dataset was included in the training set, and the gold-standard Propedia dataset was randomly split to create separate training, validation, and test sets. Specifically, 80\% of the Propedia data was allocated for training and then combined with the full PepNN training data, while the remaining 20\% was split evenly into a validation set and a held-out test set, resulting in 10\% for each. This splitting strategy ensured a balanced approach for model training and performance evaluation.

\begin{figure}[h]
  \centering
  \includegraphics[width=0.8\textwidth]{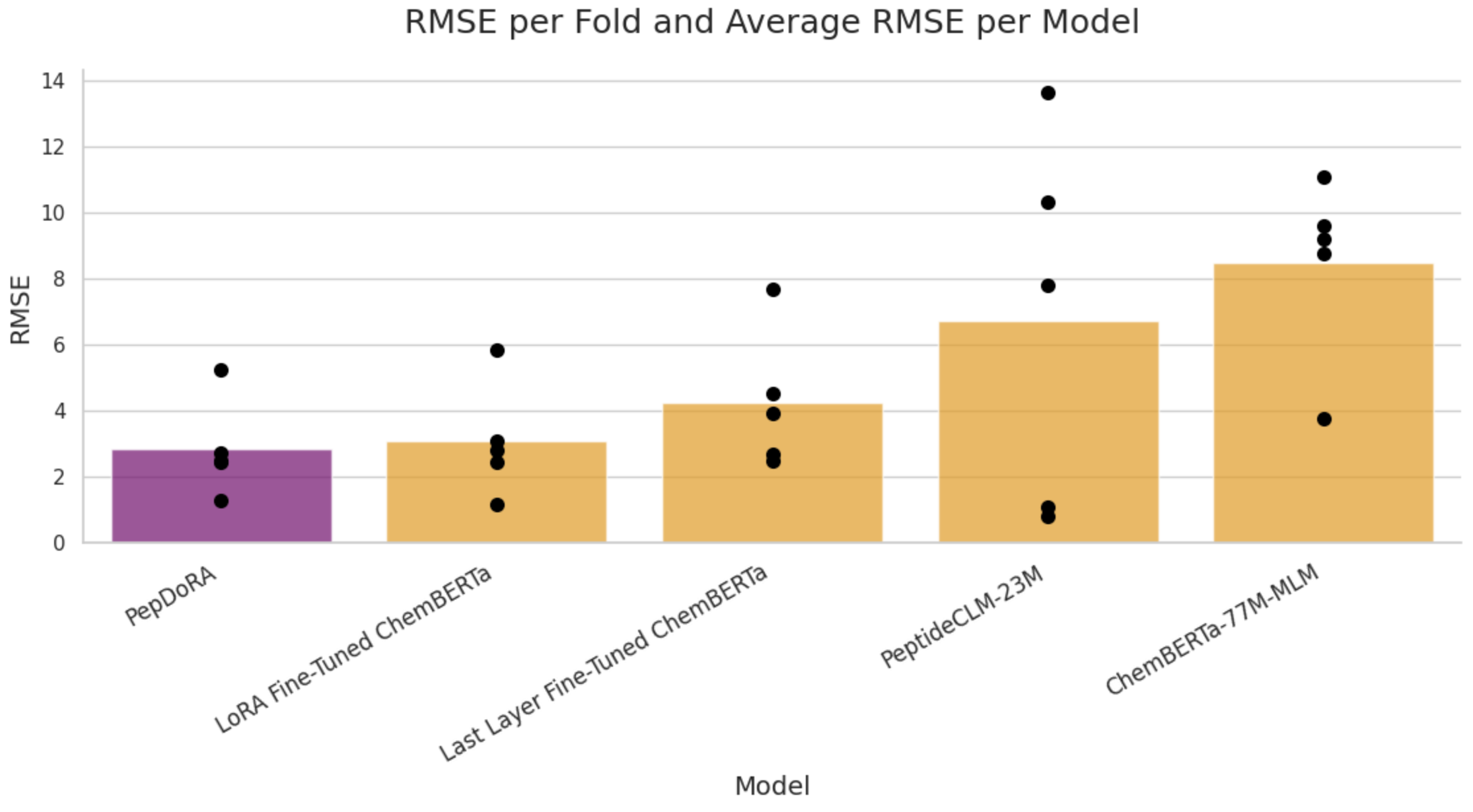}
  \caption{Comparison of membrane permeability prediction results across different model embeddings. Each bar represents the average RMSE, with individual data points indicating the RMSE values for each cross-validation fold, demonstrating the variability and consistency of each model's performance.}
  \label{Fig: membrane}
\end{figure}

\section{Results}

Using three fine-tuned ChemBERTa variants (last layer unfrozen, LoRA, and DoRA), as well as the recent PeptideCLM-23M \citep{Feller2024}, PeptideBERT \citep{Guntuboina2023}, and ChemBERTa-77M-MLM \citep{Ahmad} models, we evaluated performance on predicting the therapeutic properties of peptides, including as membrane permeability, non-fouling and hemolysis propensity, and target-specific binding. These four properties are essential in assessing the therapeutic potential of peptides, particularly those with modifications, as they collectively influence a peptide's ability to enter cells, avoid nonspecific interactions, ensure biocompatibility, interact with desired targets, and achieve clinically relevant therapeutic effects.

The ability to predict membrane permeability is crucial for determining whether peptides can effectively cross cell membranes, which is a prerequisite for intracellular peptide drug targets \citep{Yang2014, Wang2022}. We used the PAMPA dataset \citep{Siramshetty2021} for evaluating membrane permeability prediction, following the benchmark strategy outlined in the PeptideCLM paper \citep{Feller2024}. The ChemBERTa model fine-tuned via DoRA (PepDoRA) demonstrated superior performance with the lowest root mean squared error (RMSE) compared to the other models, demonstrating the model's strong adaptation capabilities to understand the chemical properties of both natural and modified peptides, making it suitable for developing therapeutic peptides capable of crossing cellular barriers (Figure \ref{Fig: membrane})

Next, we compared PepDoRA to PeptideBERT on the prediction tasks of hemolysis and non-fouling, two critical properties that influence peptide safety and efficacy \citep{Guntuboina2023}. Hemolysis refers to the breakdown of red blood cells, a property essential for determining whether a peptide can be safely used without causing toxicity to blood cells \citep{Oddo2016}, while non-fouling describes a peptide's resistance to nonspecific binding and surface adsorption, crucial for applications in biomaterials and drug delivery systems \citep{Keefe2013}. Both tasks were evaluated using pre-curated datasets restricted to unmodified peptides, as PeptideBERT is limited to encoding only wild-type amino acids without modifications \citep{Guntuboina2023}. Despite this constraint, PepDoRA demonstrated strong performance against PeptideBERT (Table \ref{tab:classification_metrics}), emphasizing its versatility to not only predict properties of modified peptides but also accurately handle linear peptide property prediction tasks. 

\begin{table}[h]\centering
\caption{Evaluation of PepDoRA and PeptideBERT on non-fouling and hemolysis classification tasks using accuracy, precision, recall and F1-score metrics.}\label{tab:classification_metrics}
\begin{tabular}{lcccc}\toprule
& \multicolumn{2}{c}{Non-fouling} & \multicolumn{2}{c}{Hemolysis} \\\cmidrule(lr){2-3} \cmidrule(lr){4-5}
Metric & PepDoRA & PeptideBERT & PepDoRA & PeptideBERT \\\midrule
Accuracy & \textbf{0.87} & \textbf{0.87} & 0.80 & \textbf{0.81} \\
Precision & 0.69 & \textbf{0.70} & \textbf{0.59} & \textbf{0.59} \\
Recall & \textbf{0.71} & 0.67 & \textbf{0.27} & 0.25 \\
F1 Score & \textbf{0.70} & 0.69 & \textbf{0.37} & 0.35 \\
\bottomrule
\end{tabular}
\end{table}

\begin{figure}[h]
  \centering
  \includegraphics[width=0.8\textwidth]{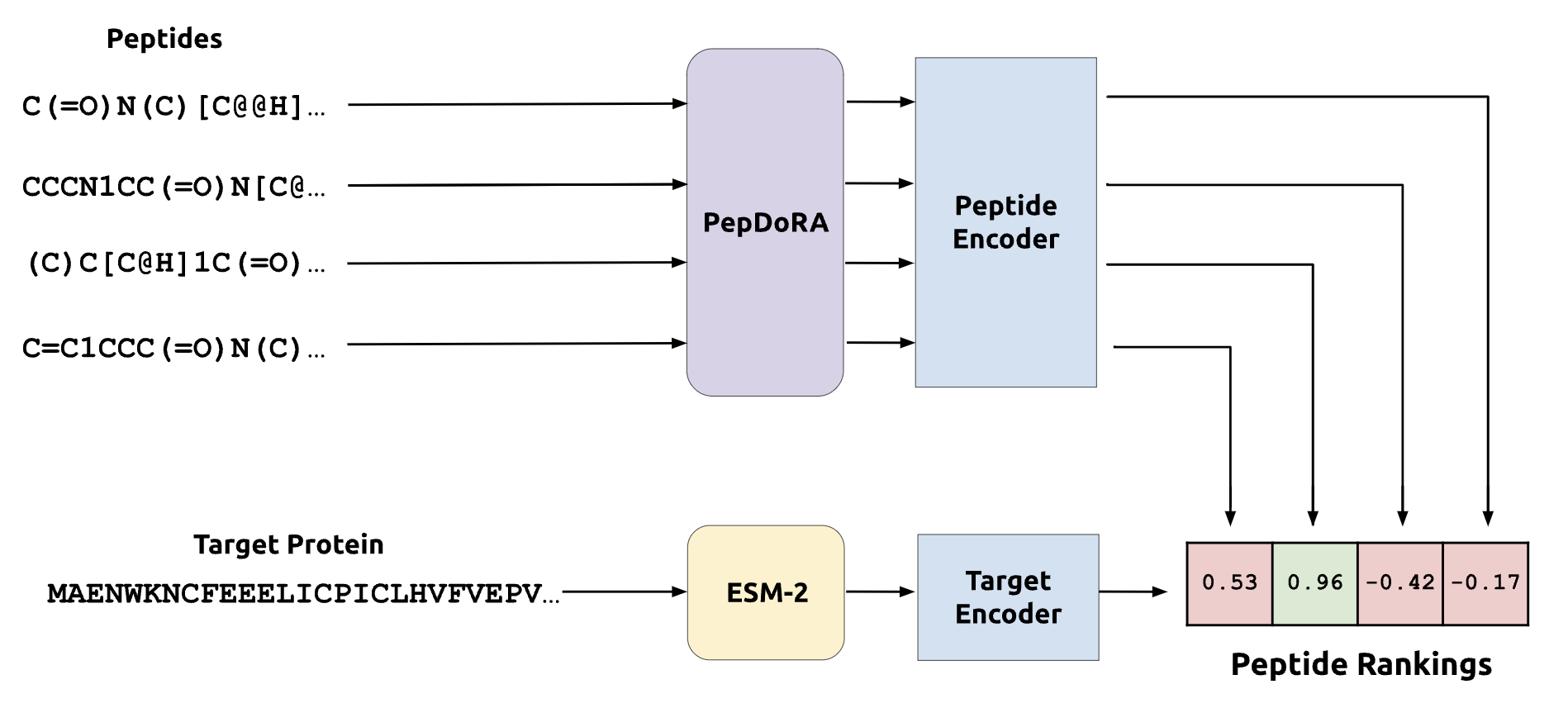}
  \caption{Schematic of the peptide-protein interaction contrastive model, illustrating learning of joint embeddings for peptide and protein pairs. The model aims to maximize cosine similarity for true binding pairs while minimizing it for non-binding pairs via a contrastive loss.}
  \label{Fig: clip}
\end{figure}

Accurately predicting binding interactions is crucial for designing peptides that can specifically target proteins, especially in cases involving modified peptides and undruggable targets. To assess the potential of PepDoRA embeddings for therapeutic biologics design, we evaluated its ability to capture peptide-protein interactions using a CLIP-based contrastive model. Here, we employed an architecture analogous to our recent PepPrCLIP architecture \citep{Bhat2023}, using peptide embeddings in combination with target ESM-2-650M protein embeddings, with the training goal of maximizing the cosine similarity between true peptide-protein pairs (Figure \ref{Fig: clip}). PepDoRA embeddings exhibited the strongest results in terms of binary accuracy, Top-1 accuracy, and Top 10\% accuracy, demonstrating its ability to represent peptides effectively and enable featurization for accurate peptide-protein mapping (Table \ref{tab:clip_metrics_comparison}). Overall, these results highlight PepDoRA's utility for designing peptide therapeutics targeting specific proteins, including difficult-to-target or intrinsically disordered proteins.

\begin{table}[h]\centering
\scriptsize
\caption{Evaluation of models on peptide-protein interaction prediction using Binary Accuracy, Top-1 Accuracy, and Top 10\% Accuracy metrics.}\label{tab:clip_metrics_comparison}
\begin{tabular}{lrrr}\toprule
Model &Binary Accuracy (\%) &Top-1 Accuracy (\%) &Top 10\% Accuracy (\%) \\\midrule
Last Layer Fine-Tuned ChemBERTa &95.5 &59.1 &81.5 \\
LoRA Fine-Tuned ChemBERTa &95.1 &60.0 &84.9 \\
DoRA Fine-Tuned ChemBERTa (PepDoRA) &\textbf{95.9} &\textbf{62.1} &\textbf{86.1} \\
PeptideCLM-23M &93.9 &47.3 &74.5 \\
ChemBERTa-77M-MLM &86.8 &27.6 &61.5 \\
\bottomrule
\end{tabular}
\end{table}

\section{Conclusion}

In this work, we introduce PepDoRA, a new language model that fine-tunes the ChemBERTa cLM to predict the functional properties of both natural and modified peptides. By leveraging the recently-described DoRA method for parameter efficient fine-tuning (PEFT), we demonstrate that ChemBERTa's robust physicochemical knowledge of small molecules can be effectively adapted to peptide-specific tasks without full model retraining. As a result, PepDoRA efficiently bridges the gap between chemical and peptide modeling, combining the strengths of cLMs with a PEFT strategy to generate optimized embeddings for peptides, particularly those with complex modifications like cyclization and unnatural amino acids. Our results show that PepDoRA performs strongly against alternative fine-tuning approaches, as well as the modified peptide-specific PeptideCLM model \citep{Feller2024} and the unmodified peptide-specific PeptideBERT model \citep{Guntuboina2023}, on multiple therapeutically-relevant tasks. 

While PepDoRA's versatile representations mark substantial progress for peptide modeling, further improvements are undoubtedly possible. Scaling the model with larger, more diverse peptide datasets and leveraging additional GPU resources will enhance its current latent space. Additionally, training on richer, more diverse peptide data from high-throughput synthesis and assays may further refine the model's representation capacity. Moving forward, we plan to integrate PepDoRA-based property predictors and our target-binding contrastive model into a conditional peptide generation algorithm, using methods such as masked discrete diffusion \citep{Goel2024}, which will be followed by experimental testing of generated candidates in our lab. As PepDoRA enables peptide representation without structural information, we plan to confirm the therapeutic efficacy of these peptides for conformationally diverse disease-related targets, especially those considered undruggable by standard small molecules \citep{Xie2023}. Ultimately, the integration of PepDoRA into our experimental workflow will advance therapeutic development, with the goal of translating these theoretical algorithms into real-world clinical solutions for previously intractable diseases.
 
\section*{Model Availability}
PepDoRA model files and weights can be freely accessed at \url{https://huggingface.co/ChatterjeeLab/PepDoRA}.


\bibliographystyle{apalike}
\bibliography{neurips_2024}  


\end{document}